\documentclass[
final            
  ]
  {aipproc}

\layoutstyle{6x9}

\begin{document}

\def\agile {\emph{AGILE}}
\def\xmm {\emph{XMM-Newton}}
\def\cha {\emph{Chandra}}
\def\flux {\mbox{erg cm$^{-2}$ s$^{-1}$}}
\def\lum {\mbox{erg s$^{-1}$}}
\def\nh {$N_{\rm H}$}

\title{On Energy-, Angular Momentum-Loss and Pulsar Spark Gaps}


\author{A. Treves}{
address={Dipartimento di Fisica, Universit\`a dell'Insubria, Via
          Valleggio 11, I-22100 Como, Italy},
altaddress={Affiliated to INAF and INFN}  
} 
 
\author{M. Pilia}{
  address={Dipartimento di Fisica, Universit\`a dell'Insubria, Via
          Valleggio 11, I-22100 Como, Italy},
altaddress={Affiliated to INAF and INFN}
}

\author{M. Lopez Moya}{
address={Universidad Complutense, E28040 Madrid, Spain}
}

\begin{abstract}
 The neutron star spin down imposes a balance between the  energy and
angular momentum  ($E/L$) losses of a pulsar. 
This translates into constraints
on the region of emission.
The $E/L$ balance may be
  cogent in discriminating among the models of $\gamma$-ray production.
Hence, models which require the release of the entire luminosity 
at the polar caps should be excluded. 
Also models where the radiative zone
 is well inside the speed of light radius have some difficulties.
It is argued that a local unbalance of $E/L$ 
should generate a global instability of the magnetosphere, possibly quenching
the emission at the polar caps.
\end{abstract}

\maketitle


\section{Introduction}
Since the seminal paper of Sturrock \cite{sturrock70}, the emission 
mechanism of pulsars is associated to spark gaps, regions where the parallel
component 
of the electric field is so large that pair production may be copious.

A key issue treated frequently in the literature is the location of the
gaps, with two favoured classes of models: a gap close to the star surface
({\it polar cap gap}), or a gap in the vicinity of the speed of light radius $r_c=
c/\omega$ ({\it outer gap}). The advantage of the former possibility is
that, close to the  
neutron star, the magnetic and electric fields are expected to be orders
of magnitude larger than at $r_c$. On the other hand, as shown originally
by Cohen \& Treves and by Holloway \cite{cohentreves72,holloway77} ,
photon release at the 
polar cap does not preserve the $E/L$ budget.
This strongly favours models with the gap located at $r_c$
(e.g. \cite{chr86}). 
However the double (or more complex) structure of pulsar pulses has generated
new interest for models where both gaps are active (see
e.g. \cite{pellizzoni09a,pellizzoni09b}). This is consistent  
with the observation that the various pulse components in general have 
different spectral shape (\cite{abdo10_vela,pilia10}).

All previous considerations are essentially of little relevance for
radio pulsars, since the fraction of the neutron star rotational energy
released in the radio band is extremely small. The problem of the location
 of the emission regions becomes of renewed interest after the observation
with the \agile\ and $Fermi$ satellites of tens of pulsars in the $\gamma$-ray
band (see \cite{pellizzoni09a,pellizzoni09b,abdo10psrcat}), since
in this band one observes a sizeable fraction of the spin-down 
 energy (10-100\%, \cite{abdo10psrcat}, see Figure 1).

\section{Consequences of the Activation of a polar cap Spark Gap}

Models could consider two spark gaps, one at the polar cap and one at the outer gap,
and the $E/L$ balance could be globally preserved. This may occur also in the case
of a continuous emission region (see e.g. the {\it slot gaps} \cite{m&h04}). 
For consistency, such models should clarify what is 
the physical mechanism which warrants that the luminosity distribution
does preserve the $E/L$ balance.

Qualitatively one can make the following argument. If part of the energy is
 released 
 at $r<r_c$, there is a local excess of angular momentum which one has to
dispose of. The only solution we can think of
is that the excess angular momentum is 
transferred to the magnetosphere, which expands to the periphery. 
The stretching of the magnetic
 field lines, may be a way of enhancing the radiation from an external gap,
again favouring the global $E/L$ balance, possibly through a relativistic
 wind.  
However the
appearence of a sizeable polar cap emission could rather induce some sort of
global instability, which, one can argue, will quench the polar cap gap.

In conclusion we expect that the activation of a polar cap gap,
if it ever occurs, should  be followed by a) a modification
 of the relativistic wind, b) enhanced power from the outer gap, c)
a global instability. The three effects may coexist or one or two may
prevail. The alternative, which is simpler and we favour, is that there is no
internal polar gap. 


\begin{figure}
  \includegraphics[height=.2\textheight]{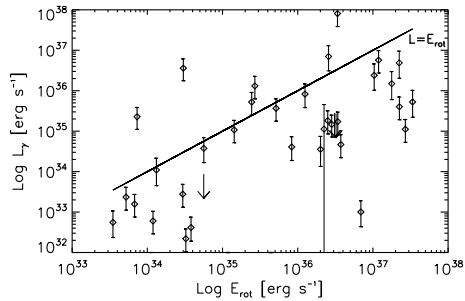}
  \caption{$\gamma$-ray luminosity of the
      pulsars observed by AGILE and Fermi as a function of the spin down
      energy loss (see also \cite{abdo10psrcat}). }
\end{figure}

\def\apj {ApJ}
\def\aap {A\&A}
\def\mnras {MNRAS}
\def\apjl {ApJL}
\def\aj {AJ}
\def\apjs {ApJS}
\def\nat {Nature}

\end{document}